\documentclass[aps,prd,twocolumn,groupedaddress,amsfonts,amssymb,amsmath,showpacs,showkeys]{revtex4-1}

\usepackage{graphicx}

\usepackage{psfrag}

\usepackage{cancel}

\begin{document}


\title{Gravitational anomalies signaling the breakdown of classical gravity}

\author{X. Hernandez\(^{1}\)}
\email[Email address: ]{xavier@astro.unam.mx}
\author{A. Jim\'enez\(^{1}\)}
\author{C. Allen\(^{1}\)}

\affiliation{\(^1\)Instituto de Astronom\'{\i}a, Universidad Nacional
                 Aut\'onoma de M\'exico, AP 70-264, Distrito Federal, 04510,
	         M\'exico \\
            }

\date{\today}


\begin{abstract}
 
Recent observations for three types of astrophysical systems severely challenge the GR plus dark matter scenario, showing a phenomenology which 
is what modified gravity theories predict. Stellar kinematics in the outskirts of globular clusters show the appearance of MOND type dynamics on 
crossing the $a_{0}$ threshold. Analysis shows a ``Tully-Fisher'' relation in these systems, a scaling of dispersion velocities with the fourth
root of their masses. Secondly, an anomaly has been found at the unexpected scales of wide binaries in the solar neighbourhood. Binary orbital 
velocities cease to fall along Keplerian expectations, and settle at a constant value, exactly on crossing the $a_{0}$ threshold. Finally, the
inferred infall velocity of the bullet cluster is inconsistent with the standard cosmological scenario, where much smaller limit encounter 
velocities appear. This stems from the escape velocity limit present in standard gravity; the ``bullet'' should not hit the ``target'' at more 
than the escape velocity of the joint system, as it very clearly did. These results are consistent with extended gravity, but would require rather
contrived explanations under GR, each. Thus, observations now put us in a situation where modifications to gravity at low acceleration scales 
cease to be a matter of choice, to now become inevitable.

\end{abstract}

\maketitle

\section{Introduction}
\label{introduction}

Over the past few years the dominant explanation for the large mass to light ratios inferred for galactic and meta-galactic 
systems, that these are embedded within massive dark matter halos, has begun to be challenged. Direct detection of the dark 
matter particles, in spite of decades of extensive and dedicated searches, remains lacking. This has led some to interpret the 
velocity dispersion measurements of stars in the local dSph galaxies (e.g. Angus 2008, Hernandez et al. 2010), the extended and 
flat rotation curves of spiral galaxies (e.g. McGaugh, Sanders \& McGaugh 2002), the large velocity dispersions of galaxies 
in clusters (e.g. Milgrom \& Sanders 2008), stellar dynamics in elliptical galaxies (e.g. Napolitano et al. 2012),
the gravitational lensing due to massive galaxies (e.g. Zhao \& Famaey 2010, Chiu et al. 2011, Mendoza et al. 2012), 
and even the cosmologically inferred matter content for the universe through CMB and structure formation physics
(e.g. Skordis et al. 2006, Halle \& Zhao 2008, Milgrom 2012), not as indirect evidence for the existence of a dominant dark matter
component, but as direct evidence for the failure of the current Newtonian and General Relativistic theories of gravity, in the 
large scale or low acceleration regimes relevant for the above situations.

Numerous alternative theories of gravity have recently appeared (e.g. TeVeS of Bekenstein 2004, and variations;  
Bruneton \& Esposito-Farese 2007, Sobouti 2007, F(R) theories e.g. Capozziello \& De Laurentis 2011, Capozziello et al. 2007, conformal 
gravity theories e.g. Mannheim \& Kazanas 1989), mostly grounded on geometrical extensions to General Relativity, and
leading to laws of gravity which in the large scale or low acceleration regime, mimic the MOdified Newtonian Dynamics (MOND) 
fitting formulas. Similarly, Mendoza et al. (2011) have explored MOND not as a modification to Newton's second law, but as a 
modified gravitational force law in the Newtonian regime, finding a good agreement with observed dynamics across galactic scales 
without requiring dark matter. In fact, recently Bernal et al. (2011) have constructed an $f(R)$ extension to general relativity 
which in the low velocity limit converges to the above approach.

A generic 
feature of all of the modified gravity schemes mentioned above is the appearance of an acceleration scale, $a_{0}$, above which 
classical gravity is recovered, and below which the dark matter mimicking regime appears. This last feature results in a general 
prediction; all systems where $a>>a_{0}$ should appear as devoid of dark matter, and all 
systems where $a<<a_{0}$ should appear as dark matter dominated, when interpreted under classical gravity.  It is interesting 
that no $a>>a_{0}$ system has ever been detected where dark matter needs to be invoked, in accordance with the former condition. 
On the other hand, the latter condition furnishes testable predictions. First, notice that for test 
particles in orbit around a $1 M_{\odot}$ star, in circular orbits of radius $s$, the acceleration is expected to drop below 
$a_{0}\approx 1.2 \times 10 ^{-10} m/s^{2}$ for $s>$7000 AU$=3.4\times10^{-2} pc$. The above provides a test for the dark matter/
modified theories of gravity debate; the relative velocities of components of binary stars with large physical separations should 
deviate from Kepler's third law under the latter interpretation.

More specifically, seen as an equivalent Newtonian force law, beyond $s \approx $7000 AU the gravitational force 
should gradually switch from the classical form of $F_{N}=GM/s^{2}$ to $F_{MG}=(G M a_{0})^{1/2}/s$, and hence the 
orbital velocity, $V^{2}/s =F$, should no longer decrease with separation, but settle at a constant value, dependent 
only on the total mass of the system through $V=(G M a_{0})^{1/4}$. That is, under modified gravity theories, 
binary stars with physical separations beyond around 7000 AU should exhibit ``flat rotation curves'' and a 
``Tully-Fisher relation'', as galactic systems in the same acceleration regime do.

In Hernandez et al. (2012a) we proposed that wide binary orbits may be used to test Newtonian gravity in the low acceleration regime.  There we applied 
this test to the binaries of two very recent catalogues containing relative velocities and separations of wide binaries.
The two catalogues are entirely independent in their approaches.

At the somewhat larger scales of globular clusters, with sizes of tens of parsecs and masses of order $10^{5} M_{\odot}$,
the central values of the stellar velocity dispersion, projected on the plane of the sky, for many Galactic globular clusters 
(GC) have been well known for decades, and are known to accurately correspond to the expectations of self-consistent dynamical models under
Newtonian gravity, e.g. King models (e.g. Binney \& Tremaine 1987, Harris 1996). Recently, a number of studies 
(e.g. Scarpa et al. 2007a, 2007b, 2010 \& 2011 and Lane et al. 2009, 2010a, 2010b \& 2011, henceforth the Scarpa et al. and Lane et al. groups respectively) 
have performed measurements of the projected velocity dispersion along the line of sight for stars in a number of Galactic GCs, but as a function of 
radius, and reaching in many cases out to radial distances larger than the half-light radii of the clusters by factors of a few.

The surprising result of the above studies has been that radially, although velocity dispersion profiles first drop along Newtonian 
expectations, after a certain radius, settle to a constant value, which varies from cluster to cluster. 
As already noted by Scarpa et al. (2011), it is suggestive of a modified gravity scenario
that the point where the velocity dispersion profiles flatten, approximately corresponds to the point where average stellar accelerations
drop below $a_{0}$. Several recent studies have shown dynamical models for self-gravitating populations of stars under MOND or
other modified gravity variants (e.g. Moffat \& Toth 2008, Haghi et al. 2009, Sollima \& Nipoti 2010, Haghi et al. 2011, Hernandez \& Jim\'enez 2012) 
which accurately reproduce not only the observed velocity dispersion profiles, but also the observed surface brightness profiles.

From the point of view of assuming Newtonian gravity to be exactly valid at all low velocity regimes, it has also been
shown that both velocity dispersion and surface brightness profiles for Galactic GCs can be self-consistently
modelled. Under this hypothesis, it is dynamical heating due to the overall Milky Way potential that is responsible for the
flattening of the velocity dispersion profiles e.g. Drukier et al. (2008), K\"{u}pper et al. (2010), Lane et al. (2010). 
The constant velocity dispersion observed at large radii merely shows the contribution of unbound stars in the process of evaporating 
into the Milky Way halo. In attempting to sort between these two contrasting scenarios, we took a fully empirical approach in Hernandez et al. (2012b). 
There, we critically examine the plausibility of both gravitational scenarios by looking through the data for other correlations which each suggest. 

For the Newtonian case, we examine the best available 
inferences for the tidal radius of each cluster at closest galacto-centric passage, and compare it to the observed point
where the velocity dispersion flattens. We found the former to generally exceed the latter by factors of 4 on average,
making the Newtonian interpretation suspect. Also, we take all the clusters which the Lane et al. group have claimed 
show no indication of a modified gravity phenomenology, based on the fact that their velocity dispersion profiles can be modelled
using Plummer profiles, and show that the fits with the generic asymptotically flat profiles we use are actually slightly better,
in all cases. 

We shall use the term MONDian to refer to any modified theory of gravity which reproduces the 
basic phenomenology of MOND in the low velocity limit for accelerations below $a_{0}$, of flat equilibrium velocities and a Tully-Fisher relation,
regardless of the details of the fundamental theory which might underlie this phenomenology.

Together with our
previous results of Hernandez \& Jimen\'ez (2012) showing that the asymptotic values of the velocity dispersion profiles 
are consistent with scaling with the fourth root of the total masses, a Tully-Fisher relation for GCs, our results support
the interpretation of the observed phenomenology as evidence for a change in regime for gravity on crossing the $a_{0}$
threshold.

A recent study reaching the same conclusions, but at a significantly distinct scale, can be found in, Lee \& Komatsu (2010) show that 
the infall velocity of the two components of the Bullet cluster, as required to
account for the hydrodynamical shock observed in the gas, is inconsistent with expectations of full cosmological simulations under
standard $\Lambda CDM$ assumptions. This has recently been confirmed at greater detail by Thompson \& Nagamine (2012), and can in fact be
seen as a failure not only of the $\Lambda CDM$ model, but of standard gravity, as the required collisional velocity is actually
larger than the escape velocity of the combined system. We note also the recent reviews by Kroupa et al. (2010), Famaey \& McGaugh (2012) and Kroupa (2012) and 
references therein, detailing a number of observations in tension with standard $\Lambda CDM$ assumptions.

\section{Dynamics of wide binaries}

The Newtonian prediction for the relative velocities of the two components of binaries having circular 
orbits, when plotted against the binary physical separation, $s$, is for a scaling of $\Delta V \propto s^{-1/2}$, essentially 
following Kepler's third law, provided the range of masses involved were narrow. 
In a relative proper motion sample however, only two components of the relative velocity appear, as velocity along the line of 
sight to the binary leads to no proper motion. Thus, orbital projection plays a part, with systems having orbital planes along 
the line of sight sometimes appearing as having no relative proper motions. A further effect comes from any degree of orbital 
ellipticity present; it is hence clear that the trend for $\Delta V \propto s^{-1/2}$ described above, will only provide an upper 
limit to the distribution of projected $\Delta V$ vs. $s$ expected in any real observed sample, even if only a narrow range of 
masses is included. One should expect a range of measured values of projected $\Delta V$ at a fixed observed projected $s$, all 
extending below the Newtonian limit, which for equal mass binaries in circular orbits gives $\Delta V_{N} =2 \left( \frac{G M}{s} \right)^{1/2}$.

Further, over time, the orbital parameters of binaries will evolve due to the effects of Galactic tidal forces and dynamical encounters 
with other stars in the field, specially in the case of wide 
binaries. To first order, one would expect little evolution for binaries tighter than the tidal limit of 1.7pc, and the 
eventual dissolution of wider systems. 
A very detailed study of all these points has recently appeared, Jiang \& Tremaine (2010). These authors numerically follow 
populations of 50,000 $1 M_{\odot}$ binaries in the Galactic environment, accounting for the evolution of the orbital parameters 
of each due to the cumulative effects of the Galactic tidal field at the Solar radius.  Also, the effects of close and long range 
encounters with other stars in the field are carefully included, to yield a present day distribution of separations and relative 
velocities for an extensive population of wide binaries, under Newtonian Gravity.

It is found that when many wide binaries cross their Jacobi radius, the two components remain fairly close by in both coordinate
and velocity space. Thus, in any real wide binary search 
a number of wide pairs with separations larger than their Jacobi radii will appear. Finally, Jiang \& Tremaine (2010) obtain the 
RMS one-dimensional relative velocity difference, $\Delta V_{1D}$, projected along a line of sight, for the entire 
populations of binaries dynamically evolved over 10 Gyr to today, as plotted against 
the projected separation on the sky for each pair. The expected Keplerian fall of $\Delta V_{1D} \propto s^{-1/2}$ for 
separations below 1.7 pc is obtained, followed by a slight rise in $\Delta V_{1D}$ as wide systems cross the Jacobi radius 
threshold. $\Delta V_{1D}$ then settles at RMS values of $ \approx 0.1 km/s$. This represents the best currently available estimate of how 
relative velocities should scale with projected separations for binary stars (both bound and in the process of dissolving in 
the Galactic tides) under Newtonian gravity. 
We see that all we need is a large sample of relative proper motion and binary separation measurements to   
test the Newtonian prediction for the RMS values of the 1 dimensional relative velocities of Jiang \& Tremaine (2010),
and to contrast the $\Delta V_{N} \propto s^{-1/2}$ and the $\Delta V_{MG} =cte.$ predictions for the upper envelope
of the $\Delta V$ vs. $s$ distributions.

\begin{figure}[!t]
\includegraphics[angle=0,scale=0.44]{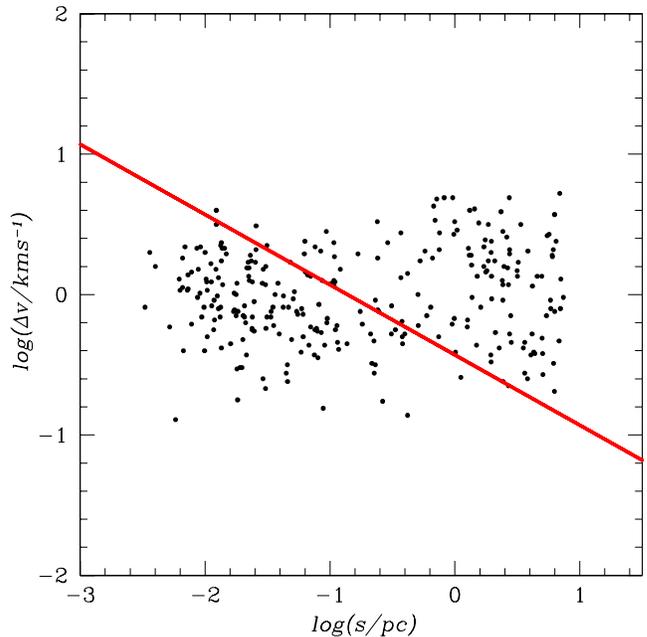}
\caption{The figure shows projected relative velocities and separations for each pair of
wide binaries from the Shaya \& Olling (2011) {\it Hipparcos} catalogue having a probability of being 
the result of chance alignment $<0.1$. The average value for the signal to noise ratio for the sample shown 
is 1.7. The upper limit shows the flat trend expected from modified gravity theories, at odds
with Kepler's third law, shown by the $s^{-1/2}$ solid line.}
\end{figure}

\begin{figure}[!t]
\includegraphics[angle=0,scale=0.44]{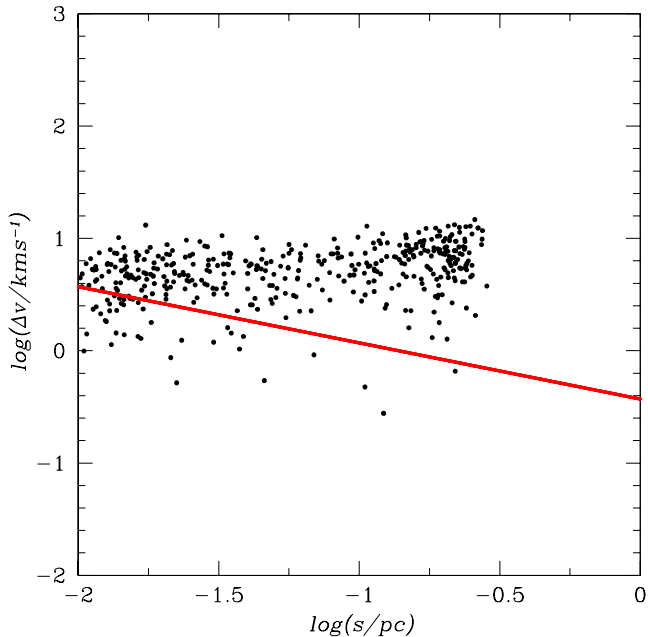}
\caption{The figure shows projected relative velocities and separations for each pair of
wide binaries from the Dhital et al. (2010) SDSS catalogue within the distance range
($225<d/pc<338$). The average value for the signal to noise ratio for the sample shown 
is 0.5. The upper limit shows the flat trend expected from modified gravity theories, at odds
with Kepler's third law, shown by the $s^{-1/2}$ solid line.}
\end{figure}

In the Shaya \& Olling (2011) catalogue wide binaries are identified by assigning a probability above chance 
alignment to the systems by carefully comparing to the underlying background (and its variations) in a 5 dimensional 
parameter space of proper motions and spatial positions. We keep only binaries with a probability 
of non-chance alignment greater than $0.9$. The binary search criteria used by the authors requires that the proposed 
binary should have no near neighbours; the projected separation between the two components is thus always many times smaller 
than the typical interstellar separation. We use the reported distances to the primaries, where errors are smallest, 
to calculate projected $\Delta V$ and projected $s$ from the measured $\Delta \mu$ and $\Delta \theta$ values reported. 
Although the use of {\it Hipparcos} measurements guarantees the best available quality in 
the data, we have also removed all binaries for which the final signal to noise ratio in 
the relative velocities was lower than $0.3$.

We are left with a sample of 280 binaries, having distances to the Sun within 
$6<d<100$ in pc. The data show a perfectly flat upper envelope in a $\Delta V$ vs. projected $s$, Hernandez \& Jim\'enez (2012).
The average signal to noise ratio for the data is 1.7, with an average error on $\Delta V$ of 0.83 $km/s$, 
which considering a $2 \sigma$ factor from the top of the distribution to the real underlying upper limit for the sample, 
results in 3 $km/s$ as our estimate of the actual physical upper limit in $\Delta V$.

The Sloan low mass wide pairs catalogue (SLoWPoKES) of Dhital et al. (2010) contains a little over 1,200 wide 
binaries with relative proper motions for each pair, distances and angular separations. Also, extreme care was 
taken to include only physical binaries, with a full galactic population model used to exclude chance alignment 
stars using galactic coordinates and galactic velocities, resulting in an estimate of fewer than 2\% of false 
positives. This yields only isolated binaries with no neighbours within many times the internal binary separation. 
Again, we use the reported distances to the primaries to calculate 
projected $\Delta V$ and projected $s$ from the measured $\Delta \mu$, $\Delta \theta$ and $d$ values reported by Dhital et al. (2010), 
to obtain a sample of 417 binaries.

The upper envelope of the distribution of 
$\Delta V$ from this catalogue does not comply with Kepler's third law. As was the case with the {\it Hipparcos} 
sample, the upper envelope describes a flat line, as expected under modified gravity 
schemes. The average signal to noise in $\Delta V$ for the Dhital et al. (2010) catalogue is
0.48, with an average error on $\Delta V$ of 12 $km/s$, which considering a $2 \sigma$ factor from the top 
of the complete distribution to the real underlying upper limit gives the same $3 km/s$ as obtained for the Shaya \& Olling (2011)
{\it Hipparcos} catalogue.

\begin{figure}[!t]
\includegraphics[angle=0,scale=0.44]{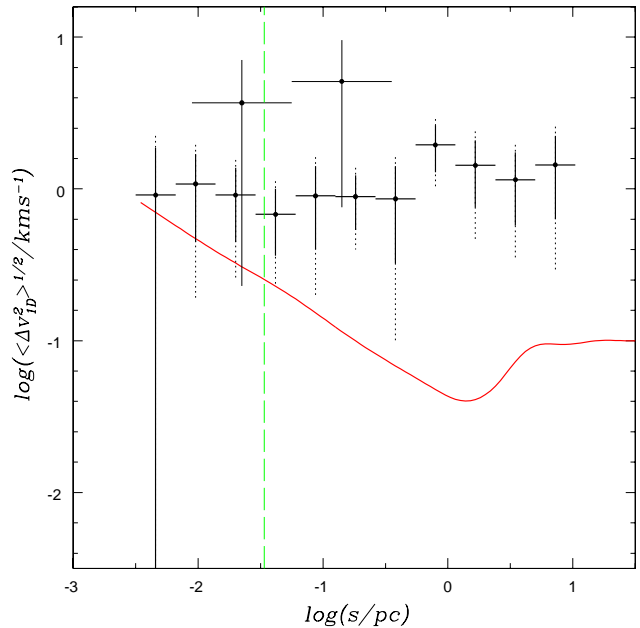}
\caption{The solid curve gives the RMS values for one dimensional projected relative velocities as a function of
projected separations, for the detailed dynamical modelling of large populations of wide binaries evolving in
the Galactic environment, taken from Jiang \& Tremaine (2010). The same quantity for the data from the catalogues analysed is
given by the points with error bars; those with narrow $log(s)$ intervals being from the {\it Hipparcos} sample of Shaya \& Olling (2011),
and those two with wide $log(s)$ intervals coming from the SDSS sample of Dhital et al. (2010).}
\end{figure}

Figure (3) shows the RMS value of the one-dimensional relative velocity difference for 
both of the samples discussed. The error bars give the error propagation on $\Delta \mu$ and $d$. We construct $\Delta V_{1D}$ 
by considering only one coordinate of the two available from the relative motion on the 
plane of the sky. Thus, each binary can furnish two $\Delta V_{1D}$ measurements, which statistically should not introduce any 
bias. Indeed, using only $\Delta \mu_{l}$ or only $\Delta \mu_{b}$ or both for each binary, yields the same mean values for the 
points shown. The small solid error bars result from considering an enlarged sample where each binary contributes two $\Delta V_
{1D}$ measurements, while the larger dotted ones come from considering each binary only once, and do not change if we consider 
only $\Delta \mu_{l}$ or only $\Delta \mu_{b}$. The series of small $log(s)$ interval data are for the {\it Hipparcos} catalogue 
of SHaya \& Olling (2011), while the two broader crosses show results for the Dhital et al. (2010) SDSS sample. 

The solid curve is the Newtonian prediction of the full Galactic evolutionary model of Jiang \& Tremaine (2010) for 
binaries, both bound and in the process of dissolving. Note that the results of this simulation deviate from Kepler's law 
for $s$ larger than the Newtonian Jacobi radius of $1.7 pc$, whereas the discrepancy with
the observed samples also occurs at much smaller separations.  Even considering the large error bars, where 
each binary contributes only one $\Delta V_{1D}$ value, we see eight points lying beyond 1$\sigma$, making the probability of 
consistency between this prediction and the observations of less than $(0.272)^{8}$=$3\times 10^{-5}$.

We obtain a constant RMS value for $\Delta V_{1D}$ of 1 $km/s$, in qualitative agreement with expectations 
from modified gravity schemes. The vertical line marks $a=a_{0}$; we see the data departing from the 
Newtonian prediction outwards of this line, and not before.
The two independent catalogues, each using different sets of selection criteria, each 
perhaps subject to its own independent systematics, are consistent with the same result, a constant horizontal upper envelope for 
the distribution of relative velocities on the plane of the sky at an intrinsic value of 3 $km/s \pm$1 $km/s$, extending over 3 
orders of magnitude in $s$, with a constant RMS $\Delta V_{1D}$ value consistent with 1 $km/s \pm$0.5 $km/s$. This supports the 
interpretation of the effect detected as the generic prediction of modified gravity theories.

\section{Outer dynamics of globular clusters}

We begin by modelling the observed projected radial velocity dispersion profiles, $\sigma_{obs}(R)$, for the globular clusters in our sample,
Hernandez et al. (2012b). As seen from the Scarpa et al. and Lane et al. data, the observed velocity dispersion profiles show a central core 
region where the velocity dispersion drops only slightly, followed by a ``Keplerian'' zone where the drop is more pronounced. 
These first two regions are in accordance with standard Newtonian King profiles, but they are then followed by a third outermost 
region where the velocity dispersion profiles cease to fall along Keplerian expectations, and settle to fixed values out to the last 
measured point. As some of us showed in Hernandez \& Jim\'enez (2012), an accurate empirical modelling for these velocity dispersion
profiles can be achieved through the function:

\begin{figure}[!t]
\includegraphics[angle=0,scale=0.44]{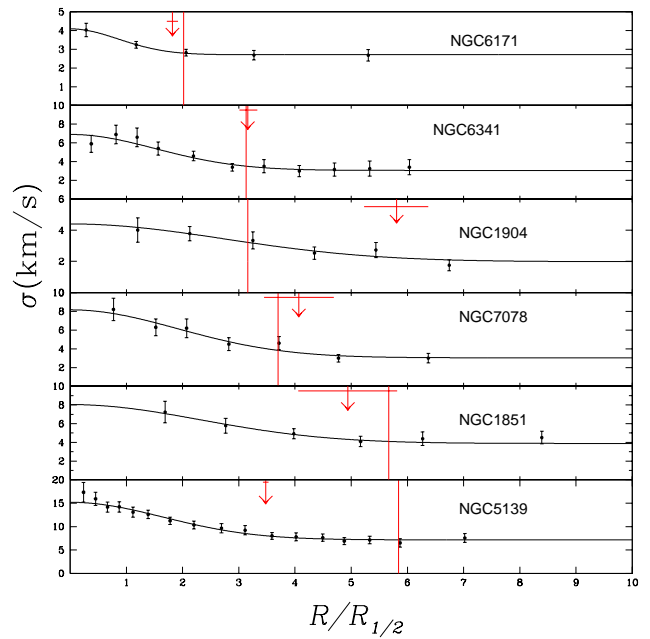}
\caption{
The figure shows the observed projected velocity dispersion profiles for six representative GCs in our sample,
points with error bars, as a function of the radial coordinate, normalised to the half-light radius of each. The solid curves 
give the maximum likelihood fits to the asymptotically flat $\sigma(R)$ model of eq.(1), seen to be accurate descriptions of the data. 
The vertical lines indicate the $a=a_{0}$ threshold, and the arrows the point where the profiles flatten, {\it a priori} 
independent features, in most cases seen to occur at approximately the same place.}
\end{figure}

\begin{equation}
\sigma(R)= \sigma_{1} e^{-(R/R_{\sigma})^{2}} + \sigma_{\infty}
\end{equation}

In the above equation $\sigma_{\infty}$ is the asymptotic value of $\sigma(R)$ at large radii, $R_{\sigma}$ a scale radius fixing how
fast the asymptotic value is approached, and $\sigma_{1}$ a normalisation constant giving $\sigma(R=0)=\sigma_{1}+\sigma_{\infty}$.

We now take the observed data points $\sigma_{obs}(R_{i})$ along with the errors associated to each data point, to determine
objectively through a maximum likelihood method the best fit values for each of the three parameters in equation (1), for each 
of the 16 observed globular clusters. The confidence intervals for each of the three parameters are then 
obtained without imposing any marginalisation. This last point allows to properly account for
any correlations between the three fitted parameters when calculating any quantity derived from combinations of them, as will
be constructed in what follows.

Taking $\sigma_{obs}(R_{i})$ data from Drukier et al. (1998), Scarpa et al. (2004), (2007a), (2007b), (2010) and (2011),
Lane et al. (2009), (2010a), (2010b) and (2011) and half-light radii, $R_{1/2}$, from integrating the surface density brightness profiles of Trager et al. 
(1995), we perform a maximum likelihood fit for all the sixteen globular clusters studied. 

Figure (4) shows the observed projected velocity dispersion profiles for 6 representative 
globular clusters from our sample, points
with error bars. The radial coordinate has been normalised to the $R_{1/2}$ radius of each of the clusters.
The continuous curves show the maximum likelihood fits for each cluster, which are clearly good representations
of the data. We can now give $R_{f}=1.5R_{\sigma}$ as an adequate empirical estimate of the radius beyond which the 
dispersion velocity profile becomes essentially flat. In terms of equation (1), which can be seen to be highly consistent 
with the observed velocity dispersion profiles, $R_{f}$ is the radius such that $\sigma(R_{f})=0.1\sigma_{1}+\sigma_{\infty}$,
a good representation of the transition to the flat behaviour, as can be checked from figure (4), where the arrows
give $R_{f}$, with the horizontal lines on the arrows showing the $1\sigma$ confidence intervals on these fitted parameters.
An empirical definition of the radius where the typical acceleration felt by stars drops below $a_{0}$ can now
be given as $R_{a}$, where:

\begin{equation}
\frac{3 \sigma(R_{a})^{2}}{R_{a}}=a_{0}.
\end{equation}

Using the above definition, we can now identify $R_{a}$ for each of the globular clusters studied. The vertical lines in Figure(5)
show $R_{a}$ for each cluster, also normalised to the half-light radius of each. In the figure, clusters have been ordered by their
$R_{a}/R_{1/2}$ values, with the smallest appearing at the top, and $R_{a}/R_{1/2}$ growing towards the bottom of the figure.

The good fits shown in the studies of the Lane et al. group to the observed velocity dispersion profiles using
Plummer models are clearly not sufficient to dismiss a modified gravity interpretation, as the asymptotically flat projected dispersion 
velocity fits of the type used for full dynamical modelling under modified gravity (Hernandez \& Jimenez 2012) actually provide even slightly
better fits to the data, Hernandez et al. (2012b).

\begin{figure}[!t]
\includegraphics[angle=0,scale=0.44]{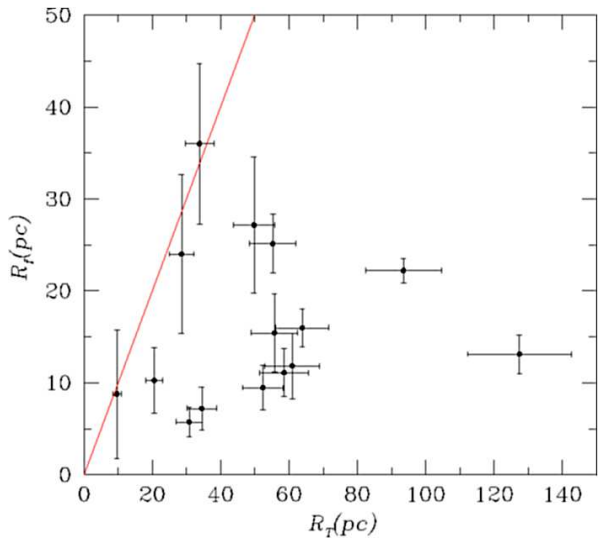}
\caption{
The figure shows the relation between the point where the velocity dispersion flattens, $R_{f}$, and the Newtonian tidal radius,
$R_{T}$, for each cluster. Even considering the large errors involved on both quantities, most points fall far to the right of the
identity line shown, making the Newtonian explanation for the flattened velocity dispersion profiles, rather suspect.}
\end{figure}

It is interesting at this point to notice a first correlation, the smaller the value of $R_{a}/R_{1/2}$, the larger the
fraction of the cluster which lies in the $a<a_{0}$ regime, and interestingly, the flatter the velocity dispersion
profile appears. At the top of the figure we see clusters where stars experience accelerations below $a_{0}$ almost at all radii, 
and it so happens, that it is only in these systems that the velocity dispersion profile appears almost flat throughout. 
Towards the bottom, we see systems where only at the outskirts accelerations fall under $a_{0}$. Over most of their extents, 
these clusters lie in the Newtonian $a>a_{0}$ regime, and indeed, it is exclusively these, that show a clear Keplerian
decline in the projected velocity dispersion profiles over most of their extents. Also, notice that $R_{f}$ and $R_{a}$ 
approximately coincide, as already previously noticed by Scarpa et al. (2007a), the flattening in the velocity dispersion 
profiles seems to appear on crossing the $a_{0}$ threshold.

In order to test the validity of the explanation for the outer flattening of the observed velocity dispersion profiles under 
Newtonian gravity, that these indicate dynamical heating due to the tides of the Milky Way system (bulge plus disk plus dark halo),
we need accurate estimates of the Newtonian tidal radii for the clusters studied.  One of us in Allen et al. (2006) and 
Allen et al. (2008) performed detailed orbital studies for 54 globular clusters for which absolute proper motions and line of sight
velocities exist. In that study, both a full 3D axisymmetric Newtonian mass model for the Milky Way and a model incorporating a galactic bar 
were used to compute precise orbits for a large sample of globular clusters, which fortunately includes
the 16 of our current study. The Galactic mass models used in those papers are fully consistent with all kinematic and structural
restrictions available. Having a full mass model, together with orbits for each globular cluster, allows the calculation of the
Newtonian tidal radius, not under any ``effective mass'' approximation, but directly through the calculation of the derivative 
of the total Galactic gravitational force, including also the evaluation of gradients in the acceleration across the extent of the 
clusters, at each point along the orbit of each studied cluster. 

The Newtonian tidal radii we take for our clusters, $R_{T}$, are actually the values which results in the largest dynamical heating 
effect upon the clusters studied, those at perigalacticon. As the distance of closest approach to the centre of the Galaxy might 
vary from passage to passage, as indeed it often does, detailed orbital integration is used to take $R_{T}$ as an average
for perigalactic passages over the last 1 Gyr.

\begin{figure}[!t]
\includegraphics[angle=0,scale=0.44]{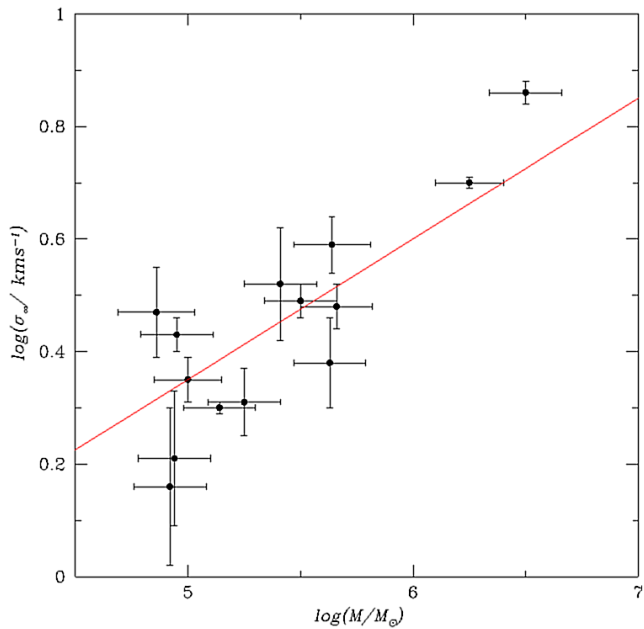}
\caption{Here we give the relation between the observed asymptotic dispersion velocity measurements, and the total mass of each
cluster. The line gives the best fit $\sigma \propto M^{1/4}$ scaling for the data.}
\end{figure}

We update the tidal radii published in Allen et al. (2006) and Allen et al. (2008),
by considering revised total masses from the integration of the observed $V$ band surface brightness profiles for our clusters
(Trager et al. 1995), and using the $V$ band stellar $M/L$ values given in McLaughlin \& van der Marel (2005) and accompanying electronic tables. 
For each individual GC, detailed single stellar population models tuned to the inferred ages and metallicities of each of the 
clusters we model were constructed in that study, using various standard population synthesis codes, and for a variety of assumed 
IMFs. In this way present stellar $M/L$ values in the $V$ band were derived, which we used. As we did not in any way use the dynamical
mass estimates of McLaughlin \& van der Marel (2005), the total masses we used are independent
of any dynamical modelling or assumption regarding the law of gravity, as they are derived through completely independent surface brightness profile
measurements and stellar population modelling. The confidence intervals in our tidal 
radii include the full range of stellar $M/L$ values given by McLaughlin \& van der Marel (2005), through considering a range
of ages, metallicities and initial mass functions consistent with the observed HR diagrams for each cluster. 

In Figure (5) we show values of $R_{f}$ for our clusters, plotted against their corresponding $R_{T}$ values, both in units of
pc. The error bars in $R_{f}$ come from full likelihood analysis in the fitting process of equation (1) to 
$\sigma_{obs}(R)$, which guarantees that confidence intervals in both of the quantities plotted are robust $1\sigma$ ranges.
The solid line shows a $R_{f}=R_{T}$ relation. It is obvious from the figure that the onset of the flat velocity dispersion regime 
occurs at radii substantially smaller than the tidal radii, for all of the globular clusters in our sample. Even under the most 
extreme accounting of the resulting errors, only three of the clusters studied are consistent with $R_{T} \approx R_{f}$ at $1\sigma$. Actually, 
the average values are closer to  $R_{T}=4 R_{f}$, with values higher than 8 appearing. One of the clusters, NGC 5024 does not appear, as 
it has values of $R_{T}=184.12$, $R_{f}=36$, which put it out of the plotted range, but consistent with the description given above.  
Given the $R^{3}$ scaling of Newtonian tidal phenomena, even a small factor of less than 2 inwards of the tidal radii, tides can be 
safely ignored, e.g. in Roche lobe overflow dynamics, the stellar interior is largely unaffected by the tidal fields, until almost 
reaching the tidal radius. It therefore appears highly unlikely under a Newtonian scheme, that Galactic tides could be responsible 
for any appreciable dynamical heating of the velocity dispersion of the studied clusters.

We note that Lane et al. (2010) and Lane et al. (2012) find that Newtonian tidal heating can explain the observed velocity dispersion 
profile of their GC sample. However, it is important to note that in Lane et al. (2010)  and Lane et al. (2012), total masses were calculated 
directly from the observed velocity dispersion observations, under the assumption that Newtonian dynamics hold. If that assumption is to be tested, 
the importance of deriving total masses through an independent method, not based on stellar dynamics, is evident.

Notice also, that most of the clusters in our sample are problematic for 
a Newtonian gravity scheme, even without the recent observations of an outer flat velocity dispersion profile. As remarked
already in Allen et al. (2006), the clusters in our sample have Newtonian tidal radii larger than the observed truncation radii
of their light distribution, the sole exceptions being Omega Cen (NGC 5139) and M92 (NGC 6341), two rather anomalous clusters.
Whereas a full dynamical modelling under an extended gravity force law of these clusters, Hernandez \& Jim\'enez (2012), naturally 
yielded an outer truncation for the light profile, under a Newtonian hypothesis, the observed truncation in the light profile of 
the clusters in our sample cannot be explained as arising from interaction with the tidal field of the Milky Way.

Furthermore, notice that we have taken $R_{T}$ at perigalacticon, where tides are at their most severe over the clusters orbit, any 
other orbital occupation averaging would result in substantially larger $R_{T}$ values. Notice also that as shown by Allen et al. (2006) 
and Allen et al. (2008), the inclusion of a realistic massive Galactic bar potential, in the case of the clusters in our sample, results 
generally in negligible changes in the resulting $R_{T}$ values, or in some cases, a slight increase in these values. Hence, even taking 
the fullest non-axisymmetric Galactic mass model under Newtonian gravity, with precise orbits derived from 3D velocity measurements for 
the clusters studied, together with total mass determinations tuned to the individual stellar populations of them, yields tidal radii as 
shown in figure (5).

As already noticed by Scarpa et al. (2011), the flattening
in the observed velocity dispersion profiles seems to appear at the point where the $a_{0}$ threshold is crossed. Here we use the much more
careful and objective modeling of the observed velocity dispersion curves of the previous section to test this point.

We see that most GCs in the sample fall within $1\sigma$ of the identity, Hernandez et al. (2012b), with about a third falling further away. 
A quantitative test of the correlation being explored is possible, since the careful modelling of the velocity dispersion profiles 
we performed naturally yields objective confidence intervals for the parameters of the fit.

We end this section with Figure (6), which shows the relation between the measured asymptotic velocity dispersion, $\sigma_{\infty}$,
and the total mass of the clusters in question. These masses represent 
the best current estimates of the stellar mass for each of the clusters in the sample, including corresponding confidence intervals.
As with all the other correlations and data presented, there is no dynamical modelling or modified gravity
assumptions going into Figure (6), merely observable quantities. We see, as already pointed out
in Hernandez \& Jim\'enez (2012), the GCs observed nicely comply with a scaling of $\sigma \propto M^{1/4}$, the Tully-Fisher
law of galactic systems ``embedded within massive dark haloes''.

The straight line shows the best fit $\sigma \propto M^{1/4}$ scaling, and actually falls only a factor of 1.3 below the modified gravity 
prediction for systems lying fully within the low acceleration regime (e.g. Hernandez \& Jim\'enez 2012), for the same value of
$a_{0}=1.2 \times 10^{-10} m/s$ used here, as calibrated through the rotation curves of galactic systems. This small offset is not surprising, 
since the GCs treated here are not fully within the $a<a_{0}$ condition, most have an inner Newtonian region encompassing a substantial 
fraction of their masses. A preliminary version of this last figure appeared already in Hernandez \& Jim\'enez (2012); we reproduce here an 
updated version using now the extended sample of clusters treated, and $\sigma_{\infty}$ values and their confidence intervals as derived through 
the careful velocity dispersion fitting procedure introduced.

To summarise, we have tested the Newtonian explanation of Galactic tides as responsible for the observed $\sigma(R)$ phenomenology,
and found it to be in tension with the observations, given the tidal radii (at perigalacticon) which the GCs in our sample present, are generally 
larger than the points where  $\sigma(R)$ flattens, on average, by factors of 4, with values higher than 8 also appearing. An explanation under a 
MONDian gravity scheme appears likely, given the clear correlations we found for the clusters in our sample, all in the expected sense.

\section{The bullet cluster and other inconcistencies of GR at galactic scales}

In going to galactic scales, the local dwarf spheroidal satellites (dSph) of the Milky Way offer an interesting test bed for
modified theories of gravity, being characterised by the highest dark matter fractions, when modelled under standard
gravity. In Hernandez et al. (2010) we showed that fully self consistent dynamical models can be constructed for these objects under 
a modified Newtonian force law, and found an interesting correlation. Assuming standard gravity, dSphs
with the oldest stellar populations show the highest dark matter fractions, something which has to be regarded as a curious coincidence.
However, under the assumption that the stars alone determine the gravitational potential of modified gravity schemes, it is natural to 
expect, indeed it is predicted, that the gravitational forces will be strongest, and deviate further from the Newtonian predictions,
for the oldest stellar populations, having higher intrinsic mass to light ratios.

More recently, Kroupa (2012) has pointed out a serious inconsistency in the standard gravity interpretation of tidal dwarf galaxies.
These small systems are formed during the interaction and merger processes of large galaxies. Within standard gravity, they are seen as
transient stellar structures displaying out of equilibrium dynamics and containing no dark matter. This last condition follows from
the large velocity dispersion values of the dark matter in the haloes of the large colliding galaxies, more than an order of
magnitude larger than what would be required to form a stable halo around the forming small tidal dwarf galaxies. However, recent
observations have confirmed that these tidal dwarf galaxies have the same dynamics as normal, stable dwarf galaxies showing
relaxed, equilibrium dynamics ``dominated by dark matter halos''. Basically, tidal dwarf galaxies lie along the same Tully-Fisher
relation as defined by normal galaxies, including more standard dwarfs. Thus, under a standard gravity plus dark matter interpretation,
we must accept as a curious coincidence that tidal dwarf galaxies, transient swirls of tidally drawn material in interacting galaxies,
have the same internal velocities as normal dwarf galaxies, equilibrium systems dominated by dark matter haloes. Under a MONDian
modified view however, the observation above is a prediction; once tides form low density ensembles of stars, these will behave
just like any other such set of stars, a standard dwarf galaxy dominated by the modified gravity regime, as internal accelerations 
are below $a_{0}$. 

At the even largest Mpc scales of the bullet cluster, a serious inconsistency in the standard gravity interpretation has been pointed out 
by Lee \& Komatsu (2010). These authors show that the encounter velocities for the two components of the bullet cluster, as determined
by the careful hydrodynamical modelling of Mastropietro \& Burkert (2008) constrained to yield the observed bow shock in the glowing x-ray gas, are so high, that they
are totally incompatible with the current standard cosmological scenario. This conclusion has recently been confirmed in much more detail by
Thompson \& Nagamine (2012), who show clearly through full standard cosmological simulations, that the probability of finding 
such a system as the observed bullet cluster, at the observed redshift, is lower than $5 \times 10^{-8}$, essentially a statement
of the impossibility of producing such a system under the standard cosmological scenario. The origin of these inconsistency is clear, 
and stems from the relation between the sound speed for the gas inside a virialized halo, and the escape velocity of these same halo.
This last velocity is the highest at which an object falling into the central more massive cluster can be expected to appear, and it is
only a factor of 2 higher than the isothermal sound speed in the gas. Thus, encounters with Mach numbers higher than two are inconsistent
with standard gravity. Within an expanding cosmological scenario, one must first overcome the expansion, making the highest Mach numbers
possible, somewhat lower than two. The strong bow shock seen in the x-ray gas of the bullet cluster however, testifies to an encounter
with a Mach number higher than 2, and hence inconsistent with standard GR and Newtonian gravity. Clearly, no amount of dark matter
can alleviate the problem, as adding dark matter will increase the impact velocity, but increase also, in the same proportion, the
sound speed in the gas before the collision, and hence leave the Mach number of the shock unchanged. One can not simply say that both
original clusters were orbiting within the higher gravitational potential of a much larger structure, as at that redshift of 0.5, in general,
no larger structures excised, indeed, none is seen in the vicinity of the cluster in question.

\section{Conclusions}

We have reviewed recent dynamical observations for wide binaries in the solar neighbourhood which show clear departures
from the Newtonian predictions on crossing the $a_{0}$ threshold, with equilibrium velocities becoming constant, as
expected under MONDian modified gravity schemes calibrated to explain rotation curves of large galaxies without the
need for any dark matter. Under a standard gravity interpretation, adding dark matter is not an option, as the scales
involved are smaller than any where dark matter is expected, and any such addition would have to be extremely fine tuned;
no dark matter for binaries having accelerations higher than $a_{0}$, and an increasing mini halo necessary
as the semi-major axis grows beyond 7000 AU.

We have shown that the velocity dispersion profiles of globular clusters in the Milky Way tend to constant velocity
values at the outskirts, at values which happen to scale precisely along the same $M^{1/4}$ relation as the galactic
Tully-Fisher relation. This is natural under any MONDian gravity scheme, but is inconsistent with standard gravity,
where tidal velocity disruptions show an inverse scaling with mass, and clear scalings with orbital parameters and
internal concentrations, none of which appear in the globular clusters studied.

At the largest scales of the colliding galaxy clusters in the bullet cluster system, it has been shown that
the standard cosmological scenario is incapable of producing the strong bow shock seen in the glowing
x-ray gas. Adding any choice amount of dark matter will not solve the problem, as the increase in the infall
velocity will be accompanied by an increase in the pre-collision sound speed of the gas, cancelling any potential
increase in the Mach number of the collision, necessary to explain the observed shock.

The recent body of observational evidence reviewed her forces a change from considering GR plus dark matter, or modified gravity, 
as equivalently plausible options; in the low velocity, $a<a_{0}$ regime, gravity does not follow the standard descriptions of 
Einstein and Newton.

\section{Acknowledgements}

Xavier Hernandez acknowledges financial assistance from UNAM DGAPA grant IN103011. Alejandra Jim\'enez 
acknowledges financial support from a CONACYT scholarship.


\begin{thebibliography}{}

\bibitem[]{} Angus, G.W., 2008, MNRAS, 387, 1481

\bibitem[]{} Allen C., Moreno E., Pichardo P., 2006, ApJ, 652, 1150

\bibitem[]{} Allen C., Moreno E., Pichardo P., 2008, ApJ, 674, 237

\bibitem[]{} Bernal T., Capozziello S., Hidalgo J. C., Mendoza S., 2011, Eur. Phys. J. C, 71, 1794

\bibitem[]{} Bekenstein J. D., 2004, Phys. Rev. D, 70, 083509

\bibitem[]{} Binney, J., Tremaine, S., 1987, Galactic Dynamics (Princeton University Press, Princeton, NJ)

\bibitem[]{} Bruneton, J. P., Esposito-Farese, G., 2007, Phys. Rev. D., 76, 124012

\bibitem[]{} Capozziello, S., Cardone, V. F., Troisi, A., 2007, MNRAS, 375, 1423

\bibitem[]{} Capozziello S., De Laurentis M., 2011, Phys. Rep. 509, 167

\bibitem[]{} Chiu, M. C., Ko, C. M., Tian, Y., Zhao, H. S., 2011, Phys. Rev. D., 83, 06352

\bibitem[]{} Dhital, S., West, A. A., Stassun, K. G., Bochanski, J. J., 2010, AJ, 139, 2566

\bibitem[]{} Drukier et al., 1998, AJ, 115, 708

\bibitem[]{} Duc, P.A., et al., 2011, MNRAS, 417, 863

\bibitem[]{} Haghi H., Baumgardt H., Kroupa P., Grebel E. K., Hilker M., Jordi K., 2009, MNRAS, 395, 1549

\bibitem[]{} Haghi H., Baumgardt H., Kroupa P., 2011, A\&A, 527, A33

\bibitem[]{} Halle, A., Zhao, H. S., Li B., 2008, ApJS, 177, 1

\bibitem[]{} Hernandez X., Mendoza S., Suarez T., Bernal T., 2010, A\&A, 514, A101

\bibitem[]{} Hernandez X., Jim\'enez M. A., Allen C., 2012a, EPJC, 72, 1884

\bibitem[]{} Hernandez X., Jim\'enez M. A., Allen C., 2012b, arXiv 1206.5024

\bibitem[]{} Hernandez X., Jim\'enez M. A., 2012, ApJ, 750, 9

\bibitem[]{} Jiang, Y. F., \& Tremaine, S., 2010, MNRAS, 401, 977 

\bibitem[]{} Kroupa, P., 2012, arXiv 1204.2546

\bibitem[]{} Lane R. R., Kiss L. L., Lewis G. F., Ibata R. A., Siebert A., Bedding T. R., Székely P., 2009, MNRAS, 400, 917

\bibitem[]{} Lane R. R. et al., 2010a, MNRAS, 406, 2732

\bibitem[]{} Lane R. R., Kiss L. L., Lewis G. F., Ibata R. A., Siebert A., Bedding T. R., Székely P., 2010b, MNRAS, 401, 2521

\bibitem[]{} Lane R. R. et al., 2011, A\&A, 530, A31

\bibitem[]{} Lane R. R., K\"{u}pper A. H. W., Heggie D. C., 2012, MNRAS, 423, 2845

\bibitem[]{} Lee J., Komatsu E., 2010, ApJ, 718, 60

\bibitem[]{} Mannheim, P. D., Kazanas, D., 1989, ApJ, 342, 635

\bibitem[]{} Mastropietro, C., Burkert, A., 2008, MNRAS, 389, 967

\bibitem[]{} McGaugh, S. S., 2011, Phys. Rev. Lett., 106, 121303

\bibitem[]{} McLaughlin D. E., van der Marel, R. P.,2005, ApJS, 161, 304

\bibitem[]{} Mendoza, S., Hernandez, X., Hidalgo, J. C., Bernal, T., 2011, MNRAS, 411, 226 

\bibitem[]{} Mendoza, S., Bernal, T., Hernandez, X., Hidalgo, J. C., Torres, L. A., 2012, arXiv 1208.6241

\bibitem[]{} Moffat J. W., Toth V. T., 2008, ApJ, 680, 1158

\bibitem[]{} Milgrom M., 1983, ApJ, 270, 365

\bibitem[]{} Milgrom M., 1994, ApJ, 429, 540

\bibitem[]{} Milgrom, M., Sanders, R. H., 2008, ApJ, 678, 131

\bibitem[]{} Milgrom, M., 2010, Phys. Rev. D., 82, 043523

\bibitem[]{} Napolitano, N. R., Capozziello, S., Romanowsky, A. J., Capaccioli, M., Tortora, C., 2012, ApJ, 748, 87

\bibitem[]{} Sanders, R. H., McGaugh, S. S., 2002, ARA\&A, 40, 263

\bibitem[]{} Scarpa R., Marconi G., Gilmozzi R., 2004, Proceedings of "Baryons in Dark Matter Halos". 
Novigrad, Croatia, 5-9 Oct 2004. Editors: R. Dettmar, U. Klein, P. Salucci. Published by SISSA, 
Proceedings of Science, http://pos.sissa.it, p.55.1

\bibitem[]{} Scarpa R., Marconi G., Gimuzzi R., Carraro G., 2007a, A\&A, 462, L9 

\bibitem[]{} Scarpa R., Marconi G., Gimuzzi R., Carraro G., 2007b, The Messenger, 128, 41

\bibitem[]{} Scarpa R., \& Falomo R., 2010, A\&A, 523, 43

\bibitem[]{} Scarpa R., Marconi G., Carraro G., Falomo R., Villanova S., 2011, A\&A, 525, A148

\bibitem[]{} Shaya, E. J., Olling, R. P., 2011, ApJS, 192, 2

\bibitem[]{} Skordis, C., Mota, D. F., Ferreira, P. G., Boehm, C., 2006, Phys. Rev. Lett. 96, 011301

\bibitem[]{} Sobouti, Y., 2007, A\&A, 464, 921

\bibitem[]{} Sollima A., Nipoti C., 2010, MNRAS, 401, 131

\bibitem[]{} Thompson R., Nagamine K., 2012, MNRAS, 419, 3560

\bibitem[]{} Trager, S.C., King, I.R., Djorgovski,S., 1995, AJ, 109,218

\bibitem[]{} Zhao H. S., Famaey B., 2010, Phys. Rev. D, 81, 087304

\end{thebibliography}
\end{document}